\documentclass[sigconf]{acmart}

\usepackage{balance}

\def\BibTeX{{\rm B\kern-.05em{\sc i\kern-.025em b}\kern-.08emT\kern-.1667em\lower.7ex\hbox{E}\kern-.125emX}}

\usepackage{booktabs} 
\usepackage{multirow}
\usepackage{mathrsfs}
\usepackage{subfigure}
\usepackage{bm}
\usepackage{threeparttable}
\usepackage{footnote}
\usepackage{enumitem}
\usepackage{url}
\usepackage{flushend}

\setcopyright{acmcopyright}
\copyrightyear{2019} 
\acmYear{2019} 
\acmConference[CIKM '19]{The 28th ACM International Conference on Information and Knowledge Management}{November 3--7, 2019}{Beijing, China}
\acmBooktitle{The 28th ACM International Conference on Information and Knowledge Management (CIKM '19), November 3--7, 2019, Beijing, China}
\acmPrice{15.00}
\acmDOI{10.1145/3357384.3357818}
\acmISBN{978-1-4503-6976-3/19/11}

\makeatletter
\g@addto@macro\normalsize{%
  \abovedisplayskip 1pt plus 1pt minus 1pt%
  \belowdisplayskip 1pt plus 1pt minus 1pt%
  \abovedisplayshortskip 1pt plus 1pt minus 1pt%
  \belowdisplayshortskip 1pt plus 1pt minus 1pt%
}
\makeatother
\begin{document}

\fancyhead{}

\title{SDM: Sequential Deep Matching Model for Online Large-scale Recommender System}

\author{Fuyu Lv$^{1}$, Taiwei Jin$^{1}$, Changlong Yu$^{2}$, Fei Sun$^{1}$, Quan Lin$^{1}$, Keping Yang$^{1}$, Wilfred Ng$^{2}$}
\affiliation{
 \institution{$^{1}$Alibaba Group, Hangzhou, China\\$^{2}$The Hong Kong University of Science and Technology, Hong Kong, China}
}
\email{{fuyu.lfy, taiwei.jtw, ofey.sf, tieyi.lq, shaoyao}@alibaba-inc.com; {cyuaq,wilfred}@cse.ust.hk}






\begin{abstract}
Capturing users' precise preferences is a fundamental problem in large-scale recommender system. Currently, item-based Collaborative Filtering (CF) methods are common matching approaches in industry. However, they are not effective to model dynamic and evolving preferences of users. In this paper, we propose a new sequential deep matching (SDM) model to capture users' dynamic preferences by combining short-term sessions and long-term behaviors. Compared with existing sequence-aware recommendation methods, we tackle the following two inherent problems in real-world applications: (1) there could exist multiple interest tendencies in one session. (2) long-term preferences may not be effectively fused with current session interests. Long-term behaviors are various and complex, hence those highly related to the short-term session should be kept for fusion. We propose to encode behavior sequences with two corresponding components: multi-head self-attention module to capture multiple types of interests and long-short term gated fusion module to incorporate long-term preferences. Successive items are recommended after matching between sequential user behavior vector and item embedding vectors. Offline experiments on real-world datasets show the superior performance of the proposed SDM. Moreover, SDM has been successfully deployed on online large-scale recommender system at Taobao and achieves improvements in terms of a range of commercial metrics.
\end{abstract}


%
%

\ccsdesc[500]{Information systems~Recommender systems}
\ccsdesc[500]{Computing methodologies~Neural networks}
\keywords{Deep Matching; Sequential Recommendation}
\maketitle
\section{Introduction}
    Large-scale recommender systems in industry are required to have both accurate prediction of users' preferences and quick response to their current need. Taobao\footnote{https://www.taobao.com/}, the largest e-commerce website in China, which supports billions of items and users, firstly retrieves a candidate set of items for a user and then applies a ranking module to generate final recommendations. In the process, the quality of candidates retrieved in the so-called matching module plays a key role in the whole system. Currently, online deployed matching models at Taobao are mainly based on item-based Collaborative Filtering (CF) methods \citeN{linden2003amazon, sarwar2001item}. However, they model static user-item interactions and do not well capture dynamic transformation in users' whole behavior sequences. Such methods usually lead to homogeneous recommendation. To accurately understand interests and preferences of users, sequential order information should be incorporated into the matching module.

    In this paper, we consider the dynamic evolution of users' interests by introducing deep sequential recommendation model instead of item-based CF in matching stage. When people begin to use online shopping services at Taobao, their behaviors accumulate to relatively long sequences. The sequences are composed of sessions. A session is a list of user behaviors that occur within a given time frame. A user usually has a clear unique shopping demand in one session \cite{quadrana2017personalizing} while his/her interest can change sharply when he/she starts a new session. Directly modeling the sequences while overlooking such the intrinsic structure would hurt the performance \cite{feng2019deep}. So we refer to the latest interaction sessions of users as short-term behaviors, other previous as long-term ones. The two parts are modeled separately to encode their inherent information which could be used to represent users' different levels of interests. Our goal is to recall top $N$ items after the user sequences as matching candidates.

    When it comes to short-term sessions modeling, methods based on recurrent neural networks (RNNs) have shown effective performance in session-based recommendation \cite{hidasi2015session}. On top of that, \citeauthor{li2017neural} \cite{li2017neural} and \citeauthor{liu2018stamp} \cite{liu2018stamp} further propose attention models to emphasize the main purpose and the effects of the last clicks respectively in a short-term session so that the models can avoid interest shift caused by users' random actions. However, they ignore that users' points of interest are multiple in a session. We observe that customers care about multiple aspects of items such as categories, brand, color, style and shop reputation, etc. Before making the final decision for the most preferred item, users compare many items repeatedly. Thus using single dot-product attention representations fails to reflect diverse interests happening in different time of purchasing. Instead, multi-head attention \cite{vaswani2017attention}, firstly proposed for machine translation tasks, allows models to jointly attend to multiple different information of different positions. The multi-head structure could naturally solve the issue of multiple interests by representing preferences from different views. So we propose our multi-interest module to augment the RNN-based sequential recommender using multi-head attention. At the same time, equipped with such self-attention, our module can represent accurate users' preferences by filtering out the causal clicks.

    Users' general preferences from long time always influence the decisions at present \cite{ying2018sequential, li2018learning, zhao2018plastic, dong2018recurrent, bai2019long}. Intuitively, if a user is a NBA basketball fan, he may view/click abundant items related to NBA stars. When he chooses to buy shoes now, sneakers of famous stars would attract him more than ordinary ones. Hence it is crucial to consider both long-term preferences and short-term behaviors. \citeauthor{ying2018sequential} \cite{ying2018sequential} and \citeauthor{li2018learning} \cite{li2018learning} both take customers' long-term preferences into account by simple combination with the current session. However, in real-world applications, customers have various and abundant shopping demands and their long-time behaviors are also complex and diverse. Stuffs related with NBA stars only take up a pretty small number of long-term behaviors.
    The long-term user preference, which is related to current short-term session, can not be significantly revealed in the overall long-term behavior representations. If we simply concatenate long- and short-term representations or sum them up over weighted attention, it is not an effective way to fuse. Information related to current short-term session in the long-term vectors should be kept.

    In our matching model, we design a gated fusion module to merge global (long-term) and local (short-term) preference features. The input to the gated fusion module is user profile embedding, long-term and short-term representation vectors. Then a gate vector is learned to control the fusion behaviors like different gates in LSTM so that the model could precisely capture interest correlation as well as users' attention to long/short-term interests. On the one hand, the most relevant information in the long-term vectors is fused with short-term vectors. On the other hand, users could have more accurate attention to long/short term interests. Unlike the scalar weights of attention-like models, the gate vector has more powerful representation ability for decision in our super complex neural networks.

    The main contributions of this paper are summarized below:
    \begin{itemize}
        \item We develop a novel sequential deep matching (SDM) model for large-scale recommender system in real-world applications by considering both short- and long-term behaviors. These two parts are modeled separately, which represent different levels of user interests.
        \item We propose to model short-term session behaviors by multi-head self-attention module to encode and capture multiple interest tendencies. A gated fusion module is used to effectively combine long-term preferences and current shopping demands, which incorporates their correlation information rather than simple combinations.
        \item Our SDM model is evaluated on two offline datasets in the real world and outperforms the other state-of-the-art methods. To demonstrate its scalability in industrial applications, we successfully deployed it on production environment of recommender system at Taobao. The SDM model has been running online effectively since December 2018 and achieves significant improvements compared to previous online system.
    \end{itemize}


\section{Related Work}\label{sec:related}
    \subsection{Deep Matching in Industry}
    To develop more effective matching models in industrial recommender system, many researchers adopt deep neural networks which have the powerful representation ability.
    Models based on Matrix Factorization (MF) \cite{koren2009matrix} try to decompose pairwise user-item implicit feedback into user and item vectors. YouTube \cite{covington2016deep} uses deep neural network to learn both embeddings of users and items. The two kinds of embeddings are generated from their corresponding features separately. The prediction is made as equivalent to search the nearest neighbors of users' vectors among all the items. Besides, \citeauthor{zhu2018learning} \cite{zhu2018learning} proposes a novel tree-based large-scale recommender system, which can provide novel items and overcome the calculation barrier of vector search.
    Recently, graph embedding based methods are applied in many industrial applications to complement or replace traditional methods. \citeauthor{wang2018billion} \cite{wang2018billion} proposes to construct an item graph from users' behavior history and then applies the state-of-the-art graph embedding methods to learn the embedding of each item. To address the cold start and sparsity problem, they incorporate side information of items to enhance the embedding procedure. \citeauthor{ying2018graph} \cite{ying2018graph} develops and deploys an effective and efficient graph convolutional network at Pinterest\footnote{https://www.pinterest.com/} to generate embeddings of nodes (items) that incorporates both graph structure as well as node feature information.
    But these models can't well take the dynamic and evolving of users' preferences into consideration. In this work, we consider this in matching stage by introducing sequential recommendation.

    \subsection{Sequence-aware Recommendation}
    Sequential recommendation aims at modeling users' preferences and predicting users' future actions such as next clicks or purchases from observed actions in a sequence manner. Previously, FPMC \cite{rendle2010factorizing} and HRM \cite{wang2015learning} model the local sequential behaviors between adjacent items in a sequence by combining Matrix Factorization and Markov Chains. Recently, deep neural networks bring powerful representation and generalization ability for recommender system.

    \citeauthor{hidasi2015session} \cite{hidasi2015session} firstly applies gated recurrent unit (GRU) to make recommendations based on users' current short sessions. Afterwards, \citeauthor{li2017neural} \cite{li2017neural} leverages attention mechanism to extract users' main purpose especially for longer sessions and achieves better results. \citeauthor{liu2018stamp} \cite{liu2018stamp} subsequently creates a novel short-term attention priority model instead of RNNs and then points out the importance of the last click in a session.
    Besides RNNs, \citeauthor{tang2018personalized} \cite{tang2018personalized} and \citeauthor{yuan2019simple} \cite{yuan2019simple} propose convolutional sequence embedding recommendation models as a solution. \citeauthor{kang2018self} \cite{kang2018self} and \citeauthor{zhang2018next} \cite{zhang2018next} use self-attention only architecture to encode user's action history. \citeauthor{tang2019towards} \cite{tang2019towards} builds a M3 model that can combine different methods above by a gating mechanism. But these methods overlook the multiple interests in one session.

    \citeauthor{chen2018sequential} \cite{chen2018sequential} introduces the memory mechanism to sequential recommender systems, which designs a user memory-augmented neural network (MANN) to express feature-level interests. As for more fine-grained user preference, \citeauthor{huang2018improving} \citeN{huang2018improving, huang2019taxonomy} use knowledge base information to enhance the semantic representation of key-value memory network called knowledge enhanced sequential recommender. However, the extra storage, manual feature design and computation of memory network of these methods cannot be accepted in industry on account of the large-scale users and items.

    It's also crucial to take customers' long-term stable preferences into consideration. \citeauthor{li2018learning} \cite{li2018learning} proposes BINN model by concatenating users' session behavior representations and stable preferences of historical purchasing behaviors. \citeauthor{ying2018sequential} \cite{ying2018sequential} comes up with a novel two-layer hierarchical attention network to recommend the next item that one user might be interested in. \citeauthor{bai2019long} \cite{bai2019long} uses multi-time scales to characterize long-short time demands and incorporate them into a hierarchical architecture. Another instance of unifying general and sequential interests is Recurrent Collaborative Filtering \cite{dong2018recurrent}, which combines RNN sequence model and matrix factorization method in a multi-task learning framework. And \citeauthor{zhao2018plastic} \cite{zhao2018plastic} does the same combination by adversarial training. Simple combinations are not effective enough to fuse short/long preferences, while multi-task and adversarial methods are not applicable in industrial applications.
    In this paper, we propose multi-head self-attention to capture multiple user interests in short-term session behaviors and use gating mechanism to incorporate long-term preferences effectively and efficiently in a real-world application.

\section{The Proposed Approach}\label{sec:approach}
    \subsection{Problem Formulation}\label{sec:formu}
    We first formulate the sequential matching problem and our solution as well as mathematical notations for variables in the deep model. Let $\mathcal{U}$ denote a set of users and $\mathcal{I}$ denote a set of items. Our model considers whether a user $u \in \mathcal{U}$ would interact with an item $i \in \mathcal{I}$ at time $t$. For user $u$, we can get his/her latest sessions by sorting the interacted items in the ascending order of time. Inspired by session-based recommendation \citeN{hidasi2015session,li2017neural}, we reformulate the new session generation rules:

    \begin{itemize}
        \item Interactions with the same session ID recorded by the backend system belong to the same one.
        \item Adjacent interactions with time difference less than 10 minutes (or longer depending on the specific scenario) are also merged into one session.
        \item Maximum length of a session is set to 50, which means a new session will begin when the session length exceeds 50.
    \end{itemize}

    Each latest session of user $u$ is regarded as the short-term behavior, namely $\mathcal{S}^u = [i_1^u, ..., i_t^u, ..., i_m^u]$, where $m$ is the length of the sequence. The long-term behaviors of $u$ that happened before $\mathcal{S}^u$ in past 7 days are denoted by $\mathcal{L}^u$.
    Based on these preliminaries, we can define our recommendation task. Given the short-term behaviors $\mathcal{S}^u$ and long-term behaviors $\mathcal{L}^u$ of user $u$, we would like to recommend items for him/her.

\begin{figure}
        \centering
        \includegraphics[scale=0.55]{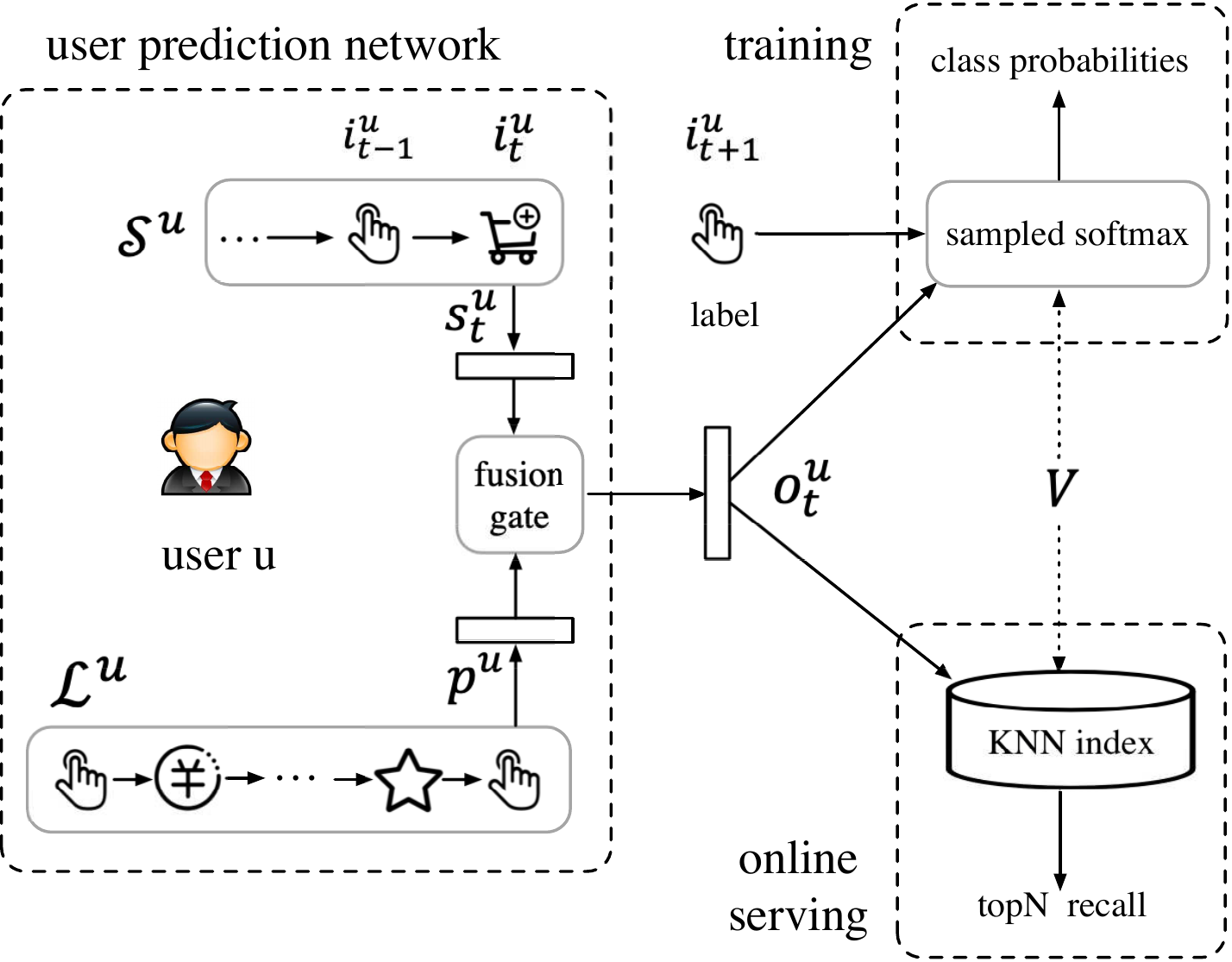}
        \caption{The general network structure of our SDM model. The user prediction network takes user's short-term session $\mathcal{S}^u = [..., i^u_{t-1}, i^u_{t}]$ and long-term behavior $\mathcal{L}^u$ as input. The target of the network is the next interacted item $i^u_{t+1}$.
        $\bm{s}^u_t$ and $\bm{p}^u$ denote short-term and long-term representations respectively. $\bm{o}^u_t$ is the predicted user behavior vector. $\bm{V}$ is the item embedding vectors.}
        \label{fig:overview}
    \end{figure}

    The general network structure is illustrated in Figure \ref{fig:overview}. Our model takes current session $\mathcal{S}^u$ and $\mathcal{L}^u$ as input. $\mathcal{S}^u$ and $\mathcal{L}^u$ are respectively encoded into short-term session representation $\bm{s}^u_t$ at time $t$ and long-term behaviors representation $\bm{p^u}$. The two kinds of representations are combined through a gated neural network. We name this module as \textit{user prediction network} that predicts user behavior vector $\bm{o}^u_t \in \mathbb{R}^{d\times1}$ from $\mathcal{S}^u$ and $\mathcal{L}^u$. Let $\bm{V} \in \mathbb{R}^{d \times |\mathcal{I}|}$ denote \textit{item embedding vectors} of $\mathcal{I}$ where $|\mathcal{I}|$ is the number of all items and $d$ is the embedding size of each vector. Our goal is to predict top $N$ item candidates at time $t+1$ based on the scores of inner product between $\bm{o}^u_t$ and each column vector $\bm{v_i}$ in $\bm{V}$
    \begin{equation}
        z_i = \mbox{score}(\bm{o}^u_t, \bm{v_i}) = {\bm{o}^u_t}^{T} \bm{v_i}
    \end{equation}
    where $\bm{v_i} \in \mathbb{R}^{d\times1}$ is the $i$th item's embedding vector.

    \begin{figure*}
        \centering
        \includegraphics[scale=0.54]{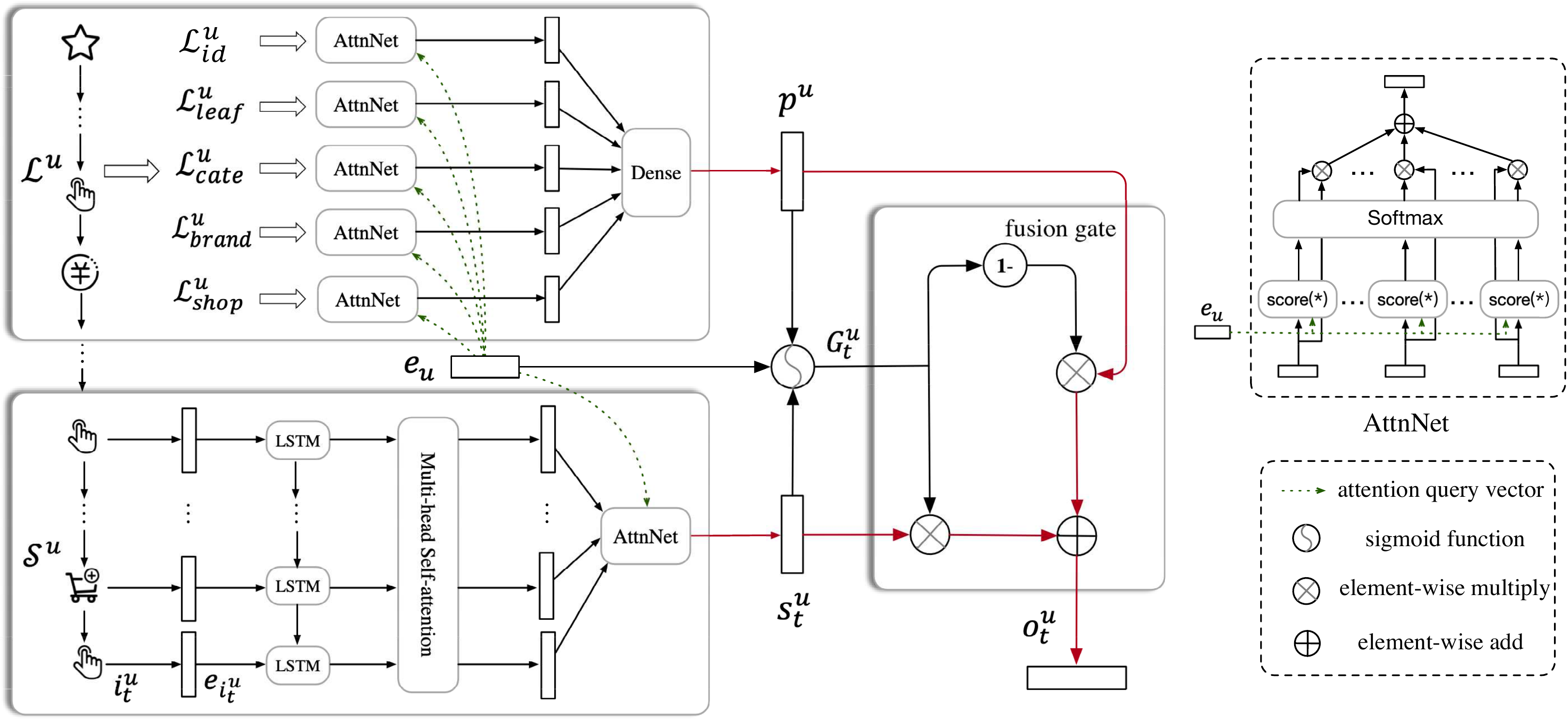}
        \caption{
        Each $i^u_t \in \mathcal{S}^u$ is embedded into a vector $\bm{e}_{i^u_t}$. Short-term representation $\bm{s}^u_t$ is encoded by LSTM and attention mechanism. We describe long-term behaviors $\mathcal{L}^u$ from various side information, i.e., item ID ($\mathcal{L}^u_{id}$), first level category ($\mathcal{L}^u_{cate}$), leaf category ($\mathcal{L}^u_{leaf}$), brand ($\mathcal{L}^u_{brand}$) and shop ($\mathcal{L}^u_{shop}$).
        Long-term representation $\bm{p}^u$ is encoded through attention and dense fully-connected networks. $\bm{s}^u_t$ and $\bm{p}^u$ are fused into user behavior vector $\bm{o}^u_t$ through a gate vector $\bm{G}^u_t$. $\bm{e}^u$ is user profile embedding.}
        \label{fig:components}
    \end{figure*}

    \subsection{Training and Online Serving}\label{sec:overview}
    During training process, the positive label at time $t$ is the next interacted item $i_{t+1}^u$. Negative labels are sampled from $\mathcal{I}$ excluding $i_{t+1}^u$ by the log-uniform sampler considering the large amount of items in a real-world application. Then the prediction class probabilities are made by a softmax layer. This is called sampled-softmax \cite{jean2014using} and we apply cross-entropy as loss function
    \begin{equation}
    \begin{split}
        \hat{\bm{y}} = \mbox{softmax}(\bm{z})\\
        L(\hat{\bm{y}}) = -\sum_{i \in \mathcal{K}} y_i \log(\hat{y}_i)
    \end{split}
    \end{equation}
    where $\mathcal{K}$ is the sampled subset of $\mathcal{I}$ including positive and negative labels, $\bm{z} = [z_1, ..., z_{|\mathcal{K}|}]$ is the inner product scores between $\bm{o}^u_t$ and each $\bm{v_i}$ ($i \in \mathcal{K}$), $\hat{\bm{y}} = [\hat{y}_1, ..., \hat{y}_{|\mathcal{K}|}]$ is the prediction probability distribution over each sampled item, and $y_i$ is the truly probability distribution of item $i$.

    We deploy the model on our online recommender system. The item embedding vectors $\bm{V}$ are imported into an efficient $K$-Nearest-Neighborhood (KNN) similarity search system \cite{JDH17}. Meanwhile, the user prediction network is deployed on a high-performance real-time inference system of machine learning. This kind of architecture follows the YouTube for video recommendation \cite{covington2016deep}.
    When customers use our online service, they will interact with lots of items and their feedback about the items will be recorded by the backend system. These information will be processed and then stored in database as users' behavior logs. Useful information from massive logs are extracted to be constructed into structured data that our model requires.
    At time $t$, customer's historical behaviors ($\mathcal{S}^u$ and $\mathcal{L}^u$) are fed into the inference system. Then the user behavior vector $\bm{o}^u_t$ is predicted. The KNN search system retrieves the most similar items with $\bm{o}^u_t$ according to their inner products. Top $N$ items are then recommended. Now we elaborate on how $\mathcal{S}^u$ and $\mathcal{L}^u$ are encoded in the network and how the two representations are fused as illustrated in Figure \ref{fig:components}.

    \subsection{Input Embedding with Side Information}\label{sec:input}
    In Taobao's recommendation scenario, customers not only focus on a specific item itself, but also concern about the brand, shop and price, etc. For example, some people tend to buy items of the specific brand, and the others would like to buy items from shops that they trust in. Furthermore, due to the sparsity caused by the large-scale online items in industry, encoding items only by item ID feature level is far from satisfaction. So, we describe an item from different feature scales, i.e., item ID, leaf category, first level category, brand and shop, which are denoted as side information set $\mathcal{F}$.
    Each input item, $i_t^u \in \mathcal{S}^u$, is represented as a dense vector $\bm{e}_{i^u_t} \in \mathbb{R}^{d \times 1}$ transformed by the embedding layer so that they can be fed into the deep neural network directly
    \begin{equation}
        \bm{e}_{i^u_t} = \mbox{concat}(\{\bm{e}^{f}_{i}|f \in \mathcal{F}\})
    \end{equation}
    where $\bm{e}^{f}_i = {\bm{W}^{f} \bm{x}_i^{f}} \in \mathbb{R}^{d_{f} \times 1}$ is item $i$'s input embedding of feature $f$ with embedding size $d_f$. $\bm{W}^{f}$ is the feature $f$'s transformation matrix and ${\bm{x}_i^{f}}$ is a one-hot vector.

    Similarly, user profile could describe user $u$ from different feature scales, such as age, gender, and life stage.
    Input of user $u$'s profile information is represented as a dense vector $\bm{e}_u \in \mathbb{R}^{d \times 1}$
    \begin{equation}
        \bm{e}_u = \mbox{concat}(\{\bm{e}^{p}_u |p \in \mathcal{P}\})
    \end{equation}
    where $\mathcal{P}$ is profile features set and $\bm{e}^{p}_u$ is embedding of feature $p$. 

    \subsection{Recurrent Layer}
    Given the embedded short-term sequence $[\bm{e}_{i^u_1}, ..., \bm{e}_{i^u_t}]$ of $u$, to capture and characterize the global temporal dependency in the short-term sequence data, we apply Long Short Term Memory (LSTM) network as the recurrent cell following session-based recommendation \citeN{hidasi2015session, li2017neural, li2018learning}. The LSTM can be described as

    \begin{equation}
    \begin{split}
    \bm{in}^u_t & = \sigma(\bm{W}^{1}_{in} {\bm{e}_{i^u_t}} + \bm{W}^{2}_{in} \bm{h}^u_{t-1} + b_{in}) \\
    \bm{f}^u_t & = \sigma(\bm{W}^{1}_{f} {\bm{e}_{i^u_t}} + \bm{W}^{2}_{f} \bm{h}^u_{t-1} + b_f) \\
    \bm{o}^u_t & = \sigma(\bm{W}^{1}_{o} {\bm{e}_{i^u_t}} + \bm{W}^2_{o} \bm{h}^u_{t-1} + b_o) \\
    \bm{c}^u_t & = \bm{f}_t \bm{c}^u_{t-1} + \bm{in}^u_t \mbox{tanh}(\bm{W}^{1}_{c}{\bm{e}_{i^u_t}} + \bm{W}^{2}_{c} \bm{h}^u_{t-1} + b_c) \\
    \bm{h}^u_t & = \bm{o}^u_t \mbox{tanh}(\bm{c}^u_t) \\
    \end{split}
    \end{equation}

    where $\bm{in}_t^{u}$, $\bm{f}_t^{u}$, $\bm{o}_t^{u}$ represent the \textit{input}, \textit{forget} and \textit{output} gates respectively. The LSTM encodes the short-term interaction sequence of $u$ into a hidden output vector $\bm{h}^u_t \in \mathbb{R}^{d\times1}$ at time $t$, which we call sequential preference representation. $\bm{c}_t^{u}$ is the cell state vector carrying information from $\bm{h}_{t-1}^{u}$ and flows between cells. We pass $\bm{h}^u_t$ to the attention network to get higher order representation. 

    \subsection{Attention Mechanism}
    Under online shopping scenario, customers usually browse some unrelated items alternatively, which are called causal clicks. Unrelated actions would somehow influence the representation of $\bm{h}^u_t$ in the sequence. We use a self-attention network to decrease the effect of those unrelated actions.
    Attention network \citeN{bahdanau2014neural, luong2015effective, wu2016google} can aggregate various vectors into an overall presentation by assigning different weight scores to each component.

    \subsubsection{Multi-head Self-Attention} 
    Self-attention is a special case of attention mechanism, which takes the sequence itself as query, key and value vectors of $d$-dimension.
    $\bm{\hat{h}}^u_t$, the output vector after self-attention, can be aggregated from previous hidden outputs of LSTM, $\bm{X}^u =  [\bm{h}^u_1, ..., \bm{h}^u_{t}]$. 

    Users may have multiple points of interest. For instance, when $u$ is browsing a skirt, both the color and novel style would be the key factors of making decisions. Single attention network would naturally be not enough to capture multiple aspect representations. Multi-head attention allows the model to jointly attend to information from different representation subspaces at different positions \cite{vaswani2017attention} and could model user preference $\bm{\hat{h}}^u_t \in \mathbb{R}^{d\times1}$ from multiple views of interest. So we import it in our attention mechanism. The output matrix, $\bm{\hat{X}}^u = [\bm{\hat{h}}^u_1, ..., \bm{\hat{h}}^u_t]$, is calculated as
    \begin{equation}
        \bm{\hat{X}}^u = \mbox{MultiHead}(\bm{X}^u) = \bm{W}^O \mbox{concat}(\mbox{\textbf{head}}^u_1, ..., \mbox{\textbf{head}}^u_h)
    \end{equation}
    where $\bm{W}^O \in \mathbb{R}^{d \times hd_k}$ denotes the weight matrix of output linear transformation, $h$ represents the amount of heads and $d_k = \frac{1}{h}{d}$.

    In detail, each $\mbox{\textbf{head}}^u_i \in \mathbb{R}^{d_k \times t}$ represents a single latent interest
    \begin{equation}
    \mbox{\textbf{head}}^u_i = \mbox{Attention}(\bm{W}_i^Q \bm{X}^u, \bm{W}_i^K \bm{X}^u, \bm{W}_i^V \bm{X}^u)
    \end{equation}
    where $\bm{W}_i^Q, \bm{W}_i^K, \bm{W}_i^V \in \mathbb{R}^{d_k \times d}$ denote linear transformation weight matrices of query, key and value respectively.
    Let $\bm{Q}^u_i = \bm{W}_i^Q \bm{X}^u$, $\bm{K}^u_i = \bm{W}_i^K \bm{X}^u$ and $\bm{V}^u_i = \bm{W}_i^V \bm{X}^u$.
    The attention score matrix is defined as follows
    \begin{equation}\label{equ:head}
    \begin{split}
    & f(\bm{Q}^u_i, \bm{K}^u_i) = {\bm{Q}^u_i}^T {\bm{K}^u_i} \\ 
    & \bm{A}^u_i = \mbox{softmax}(f(\bm{Q}^u_i, \bm{K}^u_i))
    \end{split}
    \end{equation}
    Finally, by weighted sum pooling, we get
    \begin{equation}
    \mbox{\textbf{head}}^u_i = {\bm{V}^u_i} {\bm{A}^u_i}^T
    \end{equation}

    \subsubsection{User Attention}  
    For different users, they usually show various preferences even to similar item sets. Therefore, on top of self-attention network, we add a user attention module to mine more fine-grained personalized information, where $\bm{e}_u$ is used as the query vector attending to $\bm{\hat{X}}^u = [\bm{\hat{h}}^u_1, ..., \bm{\hat{h}}^u_t]$. The short-term behavior representation $\bm{s}^u_{t} \in \mathbb{R}^{d\times1}$ at time $t$ is calculated as
    \begin{equation}
    \begin{split}
    & \alpha_k = \frac{\exp({\bm{\hat{h}}_k^{u T} \bm{e}_u)}}{\sum_{k=1}^{t}\exp(\bm{\hat{h}}_k^{u T} \bm{e}_u)} \\
    & \bm{s}^u_{t} = \sum_{k=1}^{t} \alpha_k \bm{\hat{h}}^u_k
    \end{split}
    \end{equation}

    \subsection{Long-term Behaviors Fusion}
    From long-term view, users generally accumulate interests of different level in various dimensions. A user may often visit a group of similar shops and buy items that belong to the same category repeatedly. Therefore, we also encode long-term behaviors $\mathcal{L}^u$ from different feature scales. $\mathcal{L}^u = \{\mathcal{L}^u_{f} | f \in \mathcal{F}\}$ consists of multiple subsets: $\mathcal{L}^u_{id}$ (item ID), $\mathcal{L}^u_{leaf}$ (leaf category), $\mathcal{L}^u_{cate}$ (first level category), $\mathcal{L}^u_{shop}$ (shop) and $\mathcal{L}^u_{brand}$ (brand) as illustrated in Figure \ref{fig:components}. For example, $\mathcal{L}^u_{shop}$ contains shops that $u$ has interacted in past one week. Entries in each subset are embedded and aggregated into an overall vector through an attention-based pooling considering the quick response under online environment.

    Each ${f}^u_k \in \mathcal{L}^u_{f}$ is transformed to a dense vector $\bm{g}^u_k \in \mathbb{R}^{d\times1}$ by $\bm{W}^{f}$ as in section \ref{sec:input}. Then we use user profile embedding $\bm{e}_u$ as the query vector to calculate the attention scores and acquire the representation of $\mathcal{L}^u_{f}$ as
    \begin{equation}
    \begin{split}
    & \alpha_k = \frac{\exp({\bm{g}_k^{u T} \bm{e}_u)}}{\sum_{k=1}^{|\mathcal{L}^u_{f}|}\exp(\bm{g}_k^{u T} \bm{e}_u)} \\
    & \bm{z}^u_{f} = \sum_{k=1}^{|\mathcal{L}^u_{f}|} \alpha_k \bm{g}^u_k
    \end{split}
    \end{equation}
    $\{\bm{z}^u_{f} | f \in \mathcal{F}\}$ are concatenated and fed into a fully-connected neural network
    \begin{equation}
    \begin{split}
    & \bm{z}^u  = \mbox{concat}(\{\bm{z}^u_{f} | f \in \mathcal{F}\}) \\
    & \bm{p}^u = \mbox{tanh}(\bm{W}^p \bm{z}^u + b)
    \end{split}
    \end{equation}
    where $\bm{p}^u \in \mathbb{R}^{d\times1}$ is the long-term behavior representation.

    To combine with the short-term behaviors, we elaborately design a gated neural network that takes $\bm{e}_u$, $\bm{s}^u_t$ and $\bm{p}^u$ as inputs also shown in Figure \ref{fig:components}. A gate vector $\bm{G}^u_t \in \mathbb{R}^{d\times1}$ is used to decide contribution percentages of short- and long-term at time $t$
    \begin{equation}
    \bm{G}^u_t = \mbox{sigmoid}(\bm{W}^1 \bm{e}_u + \bm{W}^2 \bm{s}^u_t + \bm{W}^3 \bm{p}^u + b)
    \end{equation}
    The final output, i.e., user behavior vector $\bm{o^u_t} \in \mathbb{R}^{d\times1}$, is computed by
    \begin{equation}
    \bm{o^u_t} = (\bm{1} - \bm{G}^u_t) \odot \bm{p^u} + \bm{G}^u_t \odot \bm{s}^u_t
    \end{equation}
    where $\odot$ is element-wise multiplication.


\section{Experiment Setup}\label{sec:exp_setup}

    \begin{table*}
      \center
      \caption{Statistics of offline and online datasets.}
      \label{tab:data}
      \begin{tabular}{c|c|c|c|c|c|c|c|c|c}
        \hline
        Dataset                 & Data Type             & Data Split &  \#{$^a$}Users     & \#Items     & \#Records     & \#Sessions & S.len{$^b$} & L.size{$^c$} & Time Interval \\
        \hline
        \multirow{2}{*}{JD}     & \multirow{2}{*}{offline} & train   & 802,479      & 154,568     & 9,653,777     & 2,666,189  & 3.3 & 20 & 15/Mar/2018 - 8/Apr/2018 \\
                                                           \cline{3-10}
                                &                          & test      & 10,366       & 74,564      & 498,492     & 15,069     & 8.6 & 20 & 9/Apr/2018 - 15/Apr/2018 \\
        \hline
        \multirow{4}{*}{Taobao} & \multirow{2}{*}{offline} & train   & 498,633      & 2,053,798   & 45,157,298    & 7,011,385  & 6.1 & 20 & 15/Dec/2018 - 21/Dec/2018  \\
                                                           \cline{3-10}
                                &                          & test    & 13,237       & 588,306     & 1,170,401     & 13,237     & 9.2 & 20 & 22/Dec/2018    \\
        \cline{2-10}
                                & \multirow{2}{*}{online}  & train   & $3.3\times10^8$ & $1\times10^8$   & $2.1\times10^{10}$  & $2.7\times10^9$ & 7.1 & 50 & Dec/2018    \\
                                                           \cline{3-10}
                                &                          & test    & $3.3\times10^8$ & $1\times10^8$   & /      & /          & 8.1 & 50  & Dec/2018  \\
        \hline
      \end{tabular}
      \\\footnotesize{$^a$\# means the number of.}
      \footnotesize{$^b$S.len is the average length of short-term behaviors.}
      \footnotesize{$^c$L.size is the maximum size of each subset in long-term behaviors.}
    \end{table*}

    \subsection{Datasets}
    We construct an offline-online train/validation/test framework to develop our model. Models are evaluated on two offline real-world e-commerce datasets. One is a large dataset sampled from daily logs of online system on \textbf{Mobile Taobao App}. The other is from \textbf{JD}\footnote{https://www.jd.com/}.
    Our code and offline datasets are available at \textcolor{blue}{\url{https://github.com/alicogintel/SDM}}.


    \textbf{Offline Dataset of Taobao}.
    We randomly select active users who interacted more than 40 items within 8 consecutive days in December 2018. In addition, we filter users whose interactions are more than 1000 items, which we believe are spam users. Then we collect their historical interaction data, in which the first 7 days for training and the 8th day for testing. We filter out items that appear less than 5 times in the dataset. Session segmentation follows the rules in section \ref{sec:formu} and we limit the maximum size of each $\mathcal{L}^u_{f}$ to 20. During training process, we remove sessions whose length are less than 2.
    In the test stage, we select approximately 10 thousands active users in the 8th day for quick evaluation. Their first 25\% short-term sessions on the 8th day are fed into models and the remaining interactions are ground truth. Besides this, customers may browse some items more than once in a day and repeating recommendation should not be encouraged, so we remain these items only once in the test data for a user.

    \textbf{Offline Dataset of JD}. 
    Because this dataset is relatively sparser and smaller, we select user-item interaction logs of three weeks for training and one week for testing. Other data construction and cleaning process are the same as Taobao. Statistic details of the two offline datasets are listed in Table \ref{tab:data}.

    \textbf{Online Dataset}. We select the most effective offline models to deploy on Taobao's production environment. The training data is from the user-item interaction logs of Mobile Taobao App from past 7 days without sampling. The same data cleaning process is applied as offline training datasets. Scales of online users and items expand to hundred million, which can cover the most active products at Taobao, and more long-term behaviors are used. Details can also be found in Table \ref{tab:data}. The online model is updated daily as well as the corresponding item and user features. 

    \subsection{Evaluation Metrics}
    \subsubsection{Offline Evaluation}
    To evaluate the offline effectiveness of different methods,
    we use \textbf{HitRate@K}, \textbf{Precision@K}, \textbf{Recall@K} and \textbf{F1@K} metrics, which are also widely used in other related works in section \ref{sec:related}.

    HitRate@K represents the proportion of test cases ($n_{hit}$) which has the correctly recommended items in a top K position in a ranking list, defined as
    \begin{displaymath}
    \mbox{HitRate@K} = \frac{n_{hit}}{N}
    \end{displaymath}
    where $N$ denotes the number of test data. In our experiment, $K=100$ and $K=20$ are used for the tests.

    Derive the recalled set of items for a user $u$ as $P_u$ ($|P_u | = K$) and the user's ground truth set as $G_u$. Precision@K reflects how many interested items for a user in the candidates. It is calculated as
    \begin{displaymath}
    \mbox{Precision@K($u$)} = \frac{|P_u \cap G_u|}{K}
    \end{displaymath}

    Recall@K represents the ability of coverage in the user's ground truth. It's calculation is
    \begin{displaymath}
    \mbox{Recall@K($u$)} = \frac{|P_u \cap G_u|}{|G_u|}
    \end{displaymath}

    To combine precision and recall, F1@K is derived as
    \begin{displaymath}
    \mbox{F1@K($u$)} = \frac{2 \times \mbox{Precision@K($u$)} \times \mbox{Recall@K($u$)}}{\mbox{Precision@K($u$)} + \mbox{Recall@K($u$)}}
    \end{displaymath}

    \subsubsection{Online Evaluation}
    We consider the most important online metrics: \textbf{pCTR}, \textbf{pGMV} and \textbf{discovery}.

    pCTR is the Click-Through-Rate per page view where each page can recommend 20 items for a user
    \begin{displaymath}
    \text{pCTR} = \frac{\#\text{clicks}}{\#\text{pages}}
    \end{displaymath}

    pGMV is the Gross Merchandise Volume per 1,000 page views. Its calculation is in same way
    \begin{displaymath}
    \text{pGMV} = 1000 \times \frac{\#\text{pay\,amount}}{\#\text{pages}}
    \end{displaymath}

    Besides the amount of online traffic and incomes, we also consider user shopping experience. Define \textit{discovery} to measure how many novel items the recommender system can provide for a user
    \begin{displaymath}
    \text{discovery} = \frac{\#\text{new\,categories}}{\#\text{all\,categories}}
    \end{displaymath}
    where the denominator is the number of all categories that a user clicks per day and the numerator is the number of new ones in past 15 days. We take the average of all users.

    \subsection{Comparison Methods}

    \begin{table*}
        \caption{Comparisons of different models on offline datasets of Taobao and JD.}
        \label{tab:offexp}
        \begin{tabular}{l|c|c|c|c|c|c|c|c}
            \hline
            \multirow{2}{*}{Models} &  \multicolumn{4}{c|}{Taobao}              &                \multicolumn{4}{c}{JD}          \\
            \cline{2-9}
                           & HitRate@100 & Recall@100 & Precision@100 & F1@100 & HitRate@20 & Recall@20 & Precision@20 & F1@20 \\
            \hline
            Item-based CF  & 60.27\% & 3.24\% & 2.00\% & 2.43\%                 & 67.50\% & 9.08\% & 9.41\% & 8.99\% \\
            DNN            & 60.88\% & 2.85\% & 1.83\% & 2.18\%                 & 68.43\% & 8.93\% & 9.65\% & 8.98\% \\
            GRU4REC        & 65.60\% & 3.66\% & 2.30\% & 2.77\%                 & 69.44\% & 9.33\% & 9.83\% & 9.29\% \\
            NARM           & 66.97\% & 3.57\% & 2.25\% & 2.70\%                 & 70.33\% & 9.07\% & 9.58\% & 9.04\% \\
            SHAN           & 67.30\% & 3.71\% & 2.33\% & 2.80\%                 & 70.54\% & 9.42\% & 10.02\% & 9.41\% \\
            BINN           & 67.55\% & 3.49\% & 2.20\% & 2.64\%                 & 72.19\% & 9.38\% & 9.93\%  & 9.36\% \\
            \hline
            SDMMA          & 68.24\% & 3.68\% & 2.32\% & 2.79\%                   & 70.41\% & 9.21\% & 9.72\% & 9.18\% \\
            PSDMMA         & 69.43\% & 3.75\% & 2.37\% & 2.84\%                   & 71.21\% & 9.21\% & 9.78\% & 9.20\% \\
            PSDMMAL        & 70.72\% & \textbf{3.86\%} & 2.44\% & \textbf{2.93\%} & 73.25\% & 9.47\% & 10.13\% & 9.48\% \\
            PSDMMAL-N      & \textbf{73.13\%} & 3.83\% & \textbf{2.45\%} & 2.92\% & \textbf{74.33\%} & \textbf{9.68\%} & \textbf{10.42\%} & \textbf{9.72\%} \\
            \hline
            PSDMMAL-NoS    & 65.41\% & 3.38\% & 2.14\% & 2.56\%                   & 70.07\%	& 9.05\% & 9.60\% &	9.03\% \\
            \hline
        \end{tabular}
    \end{table*}

    We use the following methods to compare with our model on two offline datasets. We also include five variants of our model for ablation study.
    \begin{itemize}
        \item \textbf{Item-based CF} \cite{linden2003amazon}. It's one of the major candidate generation approaches in industry. Collaborative Filtering method generates item-item similarity matrix for recommending.
        \item \textbf{DNN} \cite{covington2016deep}. A deep neural network based recommendation approach proposed by YouTube. Vectors of videos and users are concatenated and fed into a multi-layer feed forward neural network.
        \item \textbf{GRU4REC} \cite{hidasi2015session}. \citeauthor{hidasi2015session} firstly applies recurrent neural network to solve session-based recommendation, which outperforms traditional methods significantly.
        \item \textbf{NARM} \cite{li2017neural}. It is an improved version of GRU4REC with global and local attention-based structure. A hybrid encoder with an attention mechanism to model the user's sequential behavior is explored.
        \item \textbf{SHAN} \cite{ying2018sequential}. It incorporates both users' historical stable preferences and recent shopping demands with a hierarchical attention network.
        \item \textbf{BINN} \cite{li2018learning}. BINN applies RNN-based methods to encode present consumption motivations and historical purchase behaviors. It generates the unified representation by concatenation operation.
        \item \textbf{SDMMA}. \textbf{S}equential \textbf{D}eep \textbf{M}atching with \textbf{M}ulti-head \textbf{A}tte-ntion is our multi-head self-attention enhanced model. 
        \item \textbf{PSDMMA}. \textbf{P}ersonalized \textbf{SDMMA} adds user attention module to mine fine-grained personalized information.
        \item \textbf{PSDMMAL}. \textbf{PSDMMA} combines representations of short-term sessions and \textbf{L}ong-term behaviors.
        \item \textbf{PSDMMAL-N}. Based on \textbf{PSDMMAL}, during training, we take the following $\bm{N}$ items as target classes as \citeauthor{tang2018personalized} \cite{tang2018personalized} does at the current time step. $N = 5$ in this experiment.
        \item \textbf{PSDMMAL-NoS}. \textbf{PSDMMAL} does \textbf{No}t contain the embeddings of \textbf{S}ide information in short-term sessions and long-term behaviors except for the ID feature of item and user.
    \end{itemize}

    \subsection{Implementation Details}
    We implement these deep learning models by distributed Tensorflow\footnote{https://www.tensorflow.org/}. To be fair, all of these models share the same training and testing datasets, as well as input features of items and users and other training hyper-parameters. For training, we use 5 parameter severs (PSs) and 6 GPU (Tesla P100-PCIE-16GB) workers with average 30 global steps/s on offline experiment and we use 20 PSs and 100 GPU workers with average 450 global steps/s on online experiment. Adam optimizer with learning rate 0.001 is used to update parameters and gradient clipping is adopted by scaling gradients when the norm exceeded a threshold of 5. Besides this, we use a mini-batch size of 256 and sequences with similar length are organized to be a batch. Any input feature embedding and model parameters are learned from scratch without pre-training. The learnable parameters are initialized by orthogonal initializer.
    For the recurrent neural network, we use LSTM of multiple layers with dropout (probability 0.2) and residual network \cite{merity2017regularizing} between vertical LSTM stacks. The hidden unit size of LSTM is set to 64 and 128 on offline and online experiments. Different parameter settings are set respectively, because we need to evaluate model performance efficiently on offline experiment. As for the multi-head attention structure, we set the number of heads to 4 and 8 on offline and online experiments respectively. We also use layer normalization and residual adding operation as \citeauthor{vaswani2017attention} \cite{vaswani2017attention} does. The unit size of item embedding vectors, short/long-term behavior vector and user behavior vector keep the same with the one in LSTM.
    During online serving, our model only costs about \textbf{15 milliseconds} for each matching process with \textbf{top 300 items} retrieved.

\section{Empirical Analysis}\label{sec:analy}

    \begin{figure*}
        \centering
        \includegraphics[scale=0.42]{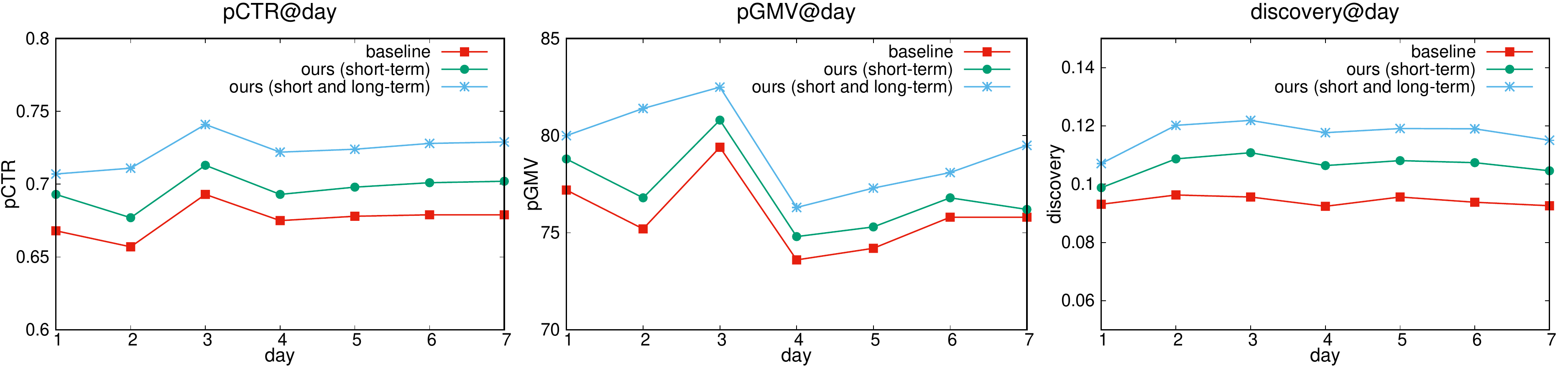}
        \caption{Online performances of our models compared with baseline in 7 days during December 2018.}
        \label{fig:online_line}
    \end{figure*}

    \subsection{Offline Results}
    Results on offline datasets of different models are shown in Table \ref{tab:offexp}. We select the best results from all the training epochs of these models. In general, deep learning based methods outperform traditional item-based CF dramatically except for YouTube-DNN. Average pooling over item sequences neglects the inherit correlation among items causing hurts on recommending quality (Recall, Precision) severely. GRU4REC and NARM consider the evolution of short-term behaviors. They perform better than original DNN models. The reason why SHAN and BINN can beat GRU4Rec in almost all metrics is that they incorporate more personalized information including long-term behaviors and user profile representation.

    Our proposed SDMMA makes use of the multi-head attention structure and has overwhelming superiority over NARM. We conduct a detailed case study in section \ref{analy:head} to further explain how multi-head attention could well capture multiple interests. By introducing user profile representation, PSDMMA strengthens the model because users of different types focus on different aspects of items. It make sense that the more accurate the short-term representation is, the more customers can find their interested items in the candidate list. But it's difficult to recommend potentially novel items for a user. More precise preferences should be inferred by considering long-term behaviors.

    PSDMMAL can beat all of the models above remarkably by taking long-term preferences into account. Different from SHAN and BINN, it applies a fusion gate to combine short- and long-term behavior representations. The gate has more powerful representation ability for decision than a hierarchical attention structure. SHAN simply applies user profile representation as query vector to decide attention weights of long-term and short-term preference representations. Our proposed PSDMMAL models the specific correlation between them. An interesting case that properly explains the design of fusion gate is shown in section \ref{analy:gate}.
    PSDMMAL-N is our best variant that takes next 5 items as target classes during training process. It can recall more diversified items that meet requirements of broader users and be more suitable for matching task of recommender system. The results of PSDMMAL-NoS show our model's performances would dramatically decrease without side information.

    \subsection{Online A/B Test}
    Currently, online matching algorithm in Taobao is a two-staged method. We define \textit{trigger items} as the latest interacted items of a user. Item-based CF firstly generates item-item similarity matrix. The trigger items recall similar items by the similarity matrix. Then these recalled items are re-ranked as matching candidates by a gradient boosting tree according to users' click and purchase logs. Such method is our online baseline and we only replace it with our model as a standard A/B test.

    We deploy PSDMMAL-N, our best sequential deep matching model, as well as the version without long-term behaviors on production environment on Mobile Taobao App. Compared with the baseline model, the quality of recommended items inferred from customers' sequence behaviors is much better than the similar ite-ms generated by item-based CF. Especially for customers who often casually browse online, our model would recommend novel items to them and attract more eyeballs to encourage potential ordering rates.
    Figure \ref{fig:online_line} shows the online results in 7 successive days during December 2018. Two sequential deep models outperformed the baseline with a large margin, where PSDMMAL-N has an overall average improvements of \textbf{7.04\%}, \textbf{4.50\%} and \textbf{24.37\%} with respect to pCTR, pGMV and discovery. Incorporating with long-term behaviors brings much more improvements. Long-term behaviors always indicate personal preferences that can affect customers' current shopping decisions. Note that our sequential matching model has been working well online since December 2018. 

    \subsection{The Effect of Multi-head Attention}\label{analy:head}

    \begin{table}
        \caption{Results of various number of heads. ($K=100$)}
        \label{tab:head}
        \begin{tabular}{c|c|c|c|c}
            \hline
            \#heads & HitRate@K & Recall@K & Precision@K & F1@k \\
            \hline
            1   & 70.00\% & 3.82\% & 2.40\% & 2.88\% \\
            2   & 70.64\% & 3.83\% & 2.41\% & 2.89\% \\
            4   & \textbf{70.72\%} & \textbf{3.86\%} & \textbf{2.44\%} & \textbf{2.93\%} \\
            8   & 70.21\% & 3.77\% & 2.37\% & 2.85\% \\
            \hline
        \end{tabular}
    \end{table}

    We explore the influence of various number of heads in our matching model. Intuitively, representation of the short-term session will get more accurate with the number of heads increasing. Table \ref{tab:head} reports the results on offline Taobao dataset. In this experiment, only the number of heads is different in PSDMMAL and the dimension of model hidden units $d$ is set to 64.

\begin{figure}
        \centering
        \includegraphics[scale=0.34]{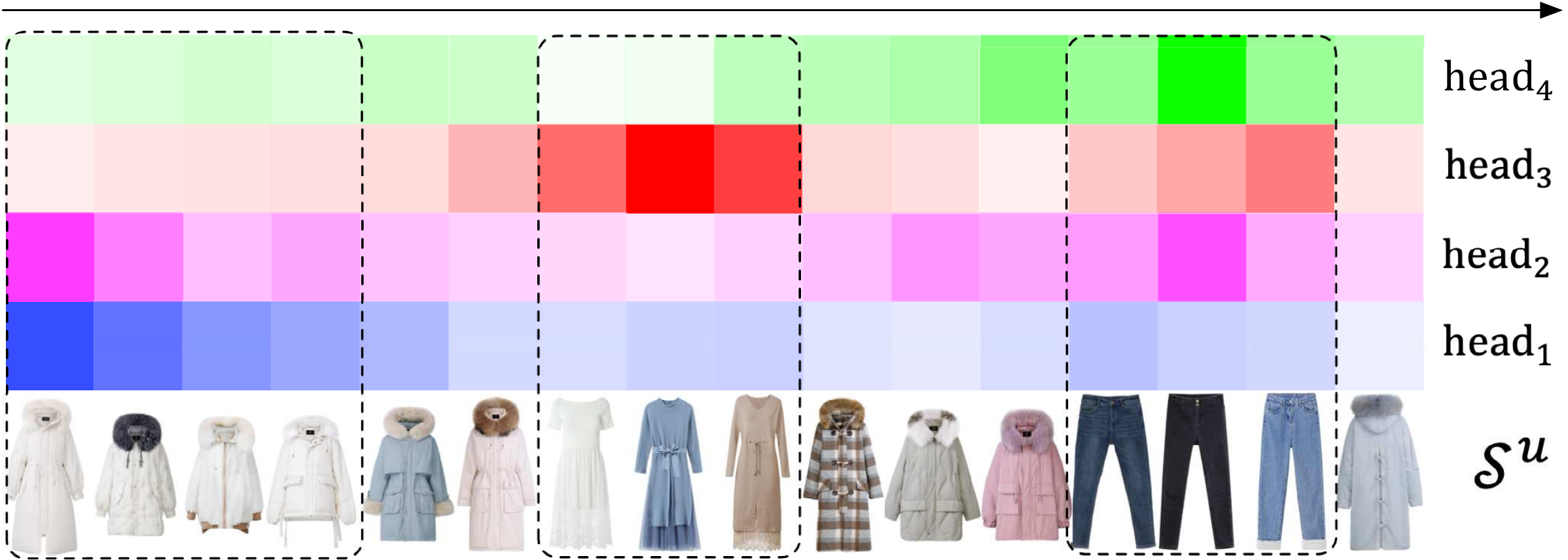}
        \caption{Visualization of attention weights (the last hidden output of LSTM as query vector to get the weights) from $\mbox{head}_1$ to $\mbox{head}_4$ over a short-term session $\mathcal{S}^u$ sampled from offline test dataset of Taobao.}
        \label{fig:attn_vis}
    \end{figure}

    We can observe that changes caused by the head number keep consistent in terms of the four metrics. When the number of heads is less than 4, the effects present positive relationship with the amount of heads. While the number of heads is greater than 4, the results become worse dramatically. We can conclude more heads are not necessarily positive because $d_{\mbox{head}_i} = \frac{64}{\#\mbox{heads}}$ would get smaller causing worse representation. In our settings, four heads can get the best results and we visualize the attention weights of the different heads over the short-term session of a user sampled from offline Taobao test dataset in Figure \ref{fig:attn_vis}.
    We choose the last hidden output $\bm{h}_t^u$ of LSTM as the query vector in multi-head attention to get the weights attending to $[\bm{h}_1^u, ..., \bm{h}_t^u]$. The weight vector is also the $t$th row vector of $\bm{A}_i^u$ in Equation \ref{equ:head}. $\mbox{head}_1$ and $\mbox{head}_2$ mainly concentrate on the first several items in the session, which are white down jackets. $\mbox{head}_3$ captures the dress interest and $\mbox{head}_4$ gives more attention to the jeans. 

    \begin{table}
        \caption{Comparisons of different fusion methods. ($K=100$)}
        \label{tab:fus}
        \begin{tabular}{c|c|c|c|c}
            \hline
            fusion   & HitRate@K & Recall@K & Precision@K & F1@K \\
            \hline
            multiply & 67.09\% & 3.42\% & 2.16\% & 2.59\% \\
            concat   & 69.74\% & 3.70\% & 2.34\% & 2.80\% \\
            add      & 70.24\% & 3.75\% & 2.37\% & 2.84\% \\
            gated    & \textbf{70.72\%} & \textbf{3.86\%} & \textbf{2.44\%} & \textbf{2.93\%} \\
            \hline
        \end{tabular}
    \end{table}

    \subsection{The Fusion Gate}\label{analy:gate}
    Element-wise multiplication, concatenation and addition operation all directly work on unfiltered long-term preference representations and ignore that a small number of preferences in long-term have strong correlation with the current short-term session. These simple combination methods take all the information from long-term preferences and naturally hurt the fusion performance shown in Figure \ref{tab:fus} tested on the Taobao offline dataset. In contrast, our proposed gated fusion network accurately captures multi-level user preferences and achieves the best results. Information highly related to current session in the long-term preference can be fused with current short-term vector.

\begin{figure}
        \centering
        \includegraphics[scale=0.285]{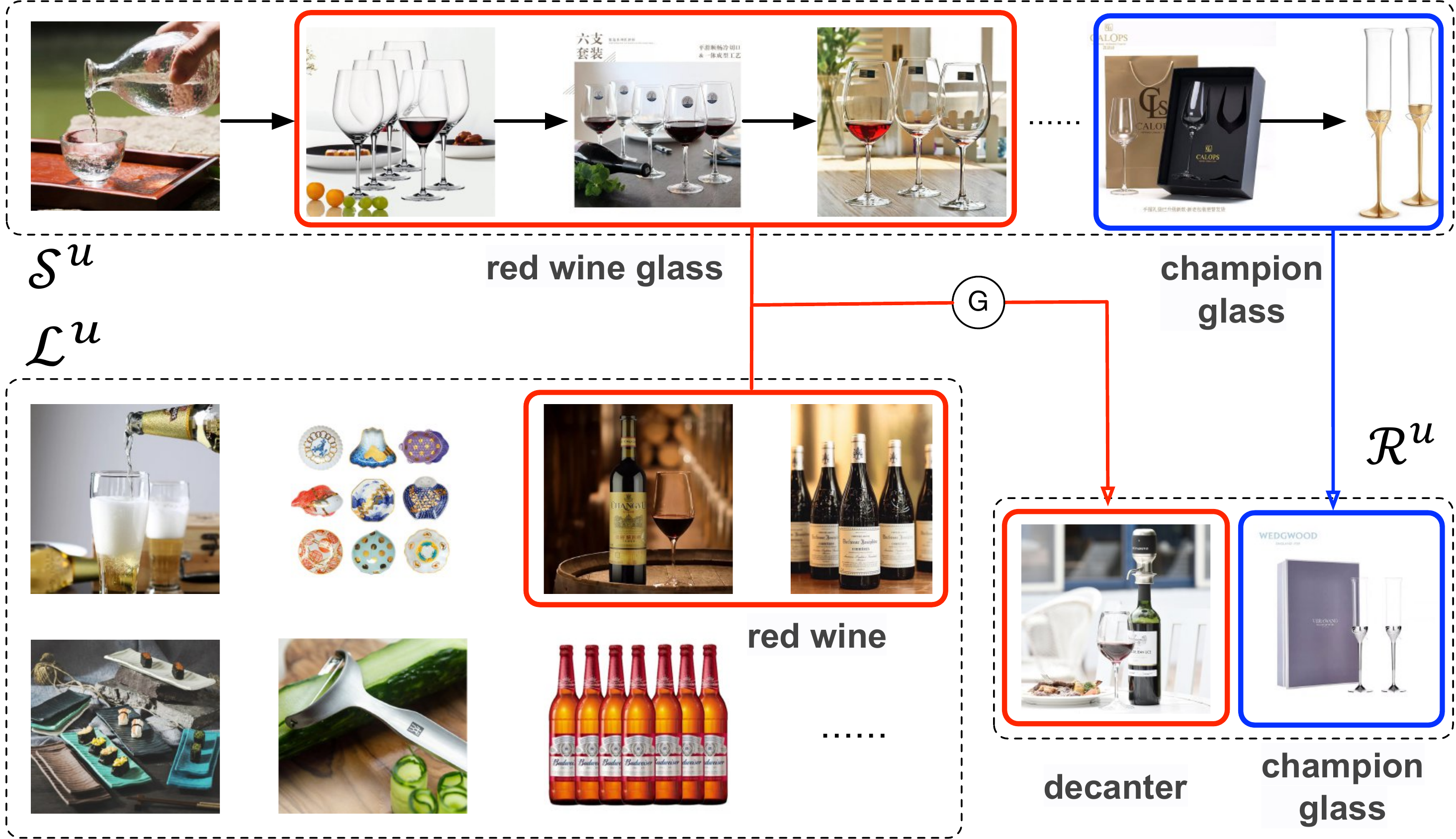}
        \caption{A short-term session $\mathcal{S}^u$ and long-term behaviors $\mathcal{L}^u$ from a sampled user on our online system. $\mathcal{R}^u$ is the set of items recommended by our model, which are also clicked by the user.}
        \label{fig:combine}
    \end{figure}

    For better explanation of the gated fusion, we use a real-world case of a sampled user at Taobao to interpret the function of gate. As shown in Figure \ref{fig:combine}, $\mathcal{R}^u$ contains items recommended by our model, which are clicked by the user simultaneously. We can see the user is browsing kinds of glasses including red wine and champion glasses. Our model can directly recommend champion glasses because they are related to the last clicks in his short-term session $\mathcal{S}^u$. It means he is more probably interested in champion glasses at present and the gate allows this information remain. Meanwhile, our gated fusion could capture the most relevant items \textbf{red wine} among his massive long-term behaviors $\mathcal{L}^u$, which also includes many irrelevant clicks such as beer, paring knife and small plate, and combine with the short-term session items \textbf{red wine glasses} to generate the recommended item \textbf{red wine decanter}. The case shows our gate module has effective and accurate fusion ability.

\section{Conclusions}\label{sec:conclusion}
    In this paper, we propose a sequential deep matching model to capture users' dynamic preferences by combining short-term sessions and long-term behaviors. We employ multi-head self-attention to capture multiple interests in short-term sessions and long-short term gated fusion network to incorporate long-term preferences. Extensive offline experiments show the effectiveness of our model. The matching model is successfully deployed on Taobao's recommender system with improvements in term of important commercial metrics.


\bibliographystyle{ACM-Reference-Format}
\bibliography{sample-bibliography}


\begin{thebibliography}{34}


\ifx \showCODEN    \undefined \def \showCODEN     #1{\unskip}     \fi
\ifx \showDOI      \undefined \def \showDOI       #1{#1}\fi
\ifx \showISBNx    \undefined \def \showISBNx     #1{\unskip}     \fi
\ifx \showISBNxiii \undefined \def \showISBNxiii  #1{\unskip}     \fi
\ifx \showISSN     \undefined \def \showISSN      #1{\unskip}     \fi
\ifx \showLCCN     \undefined \def \showLCCN      #1{\unskip}     \fi
\ifx \shownote     \undefined \def \shownote      #1{#1}          \fi
\ifx \showarticletitle \undefined \def \showarticletitle #1{#1}   \fi
\ifx \showURL      \undefined \def \showURL       {\relax}        \fi
\providecommand\bibfield[2]{#2}
\providecommand\bibinfo[2]{#2}
\providecommand\natexlab[1]{#1}
\providecommand\showeprint[2][]{arXiv:#2}

\bibitem[\protect\citeauthoryear{Bahdanau, Cho, and Bengio}{Bahdanau
  et~al\mbox{.}}{2014}]%
        {bahdanau2014neural}
\bibfield{author}{\bibinfo{person}{Dzmitry Bahdanau},
  \bibinfo{person}{Kyunghyun Cho}, {and} \bibinfo{person}{Yoshua Bengio}.}
  \bibinfo{year}{2014}\natexlab{}.
\newblock \showarticletitle{Neural machine translation by jointly learning to
  align and translate}.
\newblock \bibinfo{journal}{\emph{arXiv preprint arXiv:1409.0473}}
  (\bibinfo{year}{2014}).
\newblock


\bibitem[\protect\citeauthoryear{Bai, Du, Zhao, Wen, and Nie}{Bai
  et~al\mbox{.}}{2019}]%
        {bai2019long}
\bibfield{author}{\bibinfo{person}{Ting Bai}, \bibinfo{person}{Pan Du},
  \bibinfo{person}{Wayne~Xin Zhao}, \bibinfo{person}{Ji-Rong Wen}, {and}
  \bibinfo{person}{Jian-Yun Nie}.} \bibinfo{year}{2019}\natexlab{}.
\newblock \showarticletitle{A Long-Short Demands-Aware Model for Next-Item
  Recommendation}.
\newblock \bibinfo{journal}{\emph{arXiv preprint arXiv:1903.00066}}
  (\bibinfo{year}{2019}).
\newblock


\bibitem[\protect\citeauthoryear{Chen, Xu, Zhang, Tang, Cao, Qin, and Zha}{Chen
  et~al\mbox{.}}{2018}]%
        {chen2018sequential}
\bibfield{author}{\bibinfo{person}{Xu Chen}, \bibinfo{person}{Hongteng Xu},
  \bibinfo{person}{Yongfeng Zhang}, \bibinfo{person}{Jiaxi Tang},
  \bibinfo{person}{Yixin Cao}, \bibinfo{person}{Zheng Qin}, {and}
  \bibinfo{person}{Hongyuan Zha}.} \bibinfo{year}{2018}\natexlab{}.
\newblock \showarticletitle{Sequential recommendation with user memory
  networks}. In \bibinfo{booktitle}{\emph{WSDM}}. ACM,
  \bibinfo{pages}{108--116}.
\newblock


\bibitem[\protect\citeauthoryear{Covington, Adams, and Sargin}{Covington
  et~al\mbox{.}}{2016}]%
        {covington2016deep}
\bibfield{author}{\bibinfo{person}{Paul Covington}, \bibinfo{person}{Jay
  Adams}, {and} \bibinfo{person}{Emre Sargin}.}
  \bibinfo{year}{2016}\natexlab{}.
\newblock \showarticletitle{Deep neural networks for youtube recommendations}.
  In \bibinfo{booktitle}{\emph{RecSys}}. ACM, \bibinfo{pages}{191--198}.
\newblock


\bibitem[\protect\citeauthoryear{Dong, Zheng, Zhang, and Wang}{Dong
  et~al\mbox{.}}{2018}]%
        {dong2018recurrent}
\bibfield{author}{\bibinfo{person}{Disheng Dong}, \bibinfo{person}{Xiaolin
  Zheng}, \bibinfo{person}{Ruixun Zhang}, {and} \bibinfo{person}{Yan Wang}.}
  \bibinfo{year}{2018}\natexlab{}.
\newblock \showarticletitle{Recurrent Collaborative Filtering for Unifying
  General and Sequential Recommender}. In \bibinfo{booktitle}{\emph{IJCAI}}.
  \bibinfo{pages}{3350--3356}.
\newblock


\bibitem[\protect\citeauthoryear{Feng, Lv, Shen, Wang, Sun, Zhu, and Yang}{Feng
  et~al\mbox{.}}{2019}]%
        {feng2019deep}
\bibfield{author}{\bibinfo{person}{Yufei Feng}, \bibinfo{person}{Fuyu Lv},
  \bibinfo{person}{Weichen Shen}, \bibinfo{person}{Menghan Wang},
  \bibinfo{person}{Fei Sun}, \bibinfo{person}{Yu Zhu}, {and}
  \bibinfo{person}{Keping Yang}.} \bibinfo{year}{2019}\natexlab{}.
\newblock \showarticletitle{Deep Session Interest Network for Click-Through
  Rate Prediction}. In \bibinfo{booktitle}{\emph{IJCAI}}.
  \bibinfo{pages}{2301--2307}.
\newblock


\bibitem[\protect\citeauthoryear{Hidasi, Karatzoglou, Baltrunas, and
  Tikk}{Hidasi et~al\mbox{.}}{2015}]%
        {hidasi2015session}
\bibfield{author}{\bibinfo{person}{Bal{\'a}zs Hidasi},
  \bibinfo{person}{Alexandros Karatzoglou}, \bibinfo{person}{Linas Baltrunas},
  {and} \bibinfo{person}{Domonkos Tikk}.} \bibinfo{year}{2015}\natexlab{}.
\newblock \showarticletitle{Session-based recommendations with recurrent neural
  networks}.
\newblock \bibinfo{journal}{\emph{arXiv preprint arXiv:1511.06939}}
  (\bibinfo{year}{2015}).
\newblock


\bibitem[\protect\citeauthoryear{Huang, Ren, Zhao, He, Wen, and Dong}{Huang
  et~al\mbox{.}}{2019}]%
        {huang2019taxonomy}
\bibfield{author}{\bibinfo{person}{Jin Huang}, \bibinfo{person}{Zhaochun Ren},
  \bibinfo{person}{Wayne~Xin Zhao}, \bibinfo{person}{Gaole He},
  \bibinfo{person}{Ji-Rong Wen}, {and} \bibinfo{person}{Daxiang Dong}.}
  \bibinfo{year}{2019}\natexlab{}.
\newblock \showarticletitle{Taxonomy-aware multi-hop reasoning networks for
  sequential recommendation}. In \bibinfo{booktitle}{\emph{WSDM}}. ACM,
  \bibinfo{pages}{573--581}.
\newblock


\bibitem[\protect\citeauthoryear{Huang, Zhao, Dou, Wen, and Chang}{Huang
  et~al\mbox{.}}{2018}]%
        {huang2018improving}
\bibfield{author}{\bibinfo{person}{Jin Huang}, \bibinfo{person}{Wayne~Xin
  Zhao}, \bibinfo{person}{Hongjian Dou}, \bibinfo{person}{Ji-Rong Wen}, {and}
  \bibinfo{person}{Edward~Y Chang}.} \bibinfo{year}{2018}\natexlab{}.
\newblock \showarticletitle{Improving sequential recommendation with
  knowledge-enhanced memory networks}. In \bibinfo{booktitle}{\emph{SIGIR}}.
  ACM, \bibinfo{pages}{505--514}.
\newblock


\bibitem[\protect\citeauthoryear{Jean, Cho, Memisevic, and Bengio}{Jean
  et~al\mbox{.}}{2014}]%
        {jean2014using}
\bibfield{author}{\bibinfo{person}{S{\'e}bastien Jean},
  \bibinfo{person}{Kyunghyun Cho}, \bibinfo{person}{Roland Memisevic}, {and}
  \bibinfo{person}{Yoshua Bengio}.} \bibinfo{year}{2014}\natexlab{}.
\newblock \showarticletitle{On using very large target vocabulary for neural
  machine translation}.
\newblock \bibinfo{journal}{\emph{arXiv preprint arXiv:1412.2007}}
  (\bibinfo{year}{2014}).
\newblock


\bibitem[\protect\citeauthoryear{Johnson, Douze, and J{\'e}gou}{Johnson
  et~al\mbox{.}}{2017}]%
        {JDH17}
\bibfield{author}{\bibinfo{person}{Jeff Johnson}, \bibinfo{person}{Matthijs
  Douze}, {and} \bibinfo{person}{Herv{\'e} J{\'e}gou}.}
  \bibinfo{year}{2017}\natexlab{}.
\newblock \showarticletitle{Billion-scale similarity search with GPUs}.
\newblock \bibinfo{journal}{\emph{arXiv preprint arXiv:1702.08734}}
  (\bibinfo{year}{2017}).
\newblock


\bibitem[\protect\citeauthoryear{Kang and McAuley}{Kang and McAuley}{2018}]%
        {kang2018self}
\bibfield{author}{\bibinfo{person}{Wang-Cheng Kang} {and}
  \bibinfo{person}{Julian McAuley}.} \bibinfo{year}{2018}\natexlab{}.
\newblock \showarticletitle{Self-attentive sequential recommendation}. In
  \bibinfo{booktitle}{\emph{ICDE}}. IEEE, \bibinfo{pages}{197--206}.
\newblock


\bibitem[\protect\citeauthoryear{Koren, Bell, and Volinsky}{Koren
  et~al\mbox{.}}{2009}]%
        {koren2009matrix}
\bibfield{author}{\bibinfo{person}{Yehuda Koren}, \bibinfo{person}{Robert
  Bell}, {and} \bibinfo{person}{Chris Volinsky}.}
  \bibinfo{year}{2009}\natexlab{}.
\newblock \showarticletitle{Matrix factorization techniques for recommender
  systems}.
\newblock \bibinfo{journal}{\emph{Computer}} \bibinfo{number}{8}
  (\bibinfo{year}{2009}), \bibinfo{pages}{30--37}.
\newblock


\bibitem[\protect\citeauthoryear{Li, Ren, Chen, Ren, Lian, and Ma}{Li
  et~al\mbox{.}}{2017}]%
        {li2017neural}
\bibfield{author}{\bibinfo{person}{Jing Li}, \bibinfo{person}{Pengjie Ren},
  \bibinfo{person}{Zhumin Chen}, \bibinfo{person}{Zhaochun Ren},
  \bibinfo{person}{Tao Lian}, {and} \bibinfo{person}{Jun Ma}.}
  \bibinfo{year}{2017}\natexlab{}.
\newblock \showarticletitle{Neural attentive session-based recommendation}. In
  \bibinfo{booktitle}{\emph{CIKM}}. ACM, \bibinfo{pages}{1419--1428}.
\newblock


\bibitem[\protect\citeauthoryear{Li, Zhao, Liu, Huang, Mei, and Chen}{Li
  et~al\mbox{.}}{2018}]%
        {li2018learning}
\bibfield{author}{\bibinfo{person}{Zhi Li}, \bibinfo{person}{Hongke Zhao},
  \bibinfo{person}{Qi Liu}, \bibinfo{person}{Zhenya Huang},
  \bibinfo{person}{Tao Mei}, {and} \bibinfo{person}{Enhong Chen}.}
  \bibinfo{year}{2018}\natexlab{}.
\newblock \showarticletitle{Learning from history and present: next-item
  recommendation via discriminatively exploiting user behaviors}. In
  \bibinfo{booktitle}{\emph{KDD}}. ACM, \bibinfo{pages}{1734--1743}.
\newblock


\bibitem[\protect\citeauthoryear{Linden, Smith, and York}{Linden
  et~al\mbox{.}}{2003}]%
        {linden2003amazon}
\bibfield{author}{\bibinfo{person}{Greg Linden}, \bibinfo{person}{Brent Smith},
  {and} \bibinfo{person}{Jeremy York}.} \bibinfo{year}{2003}\natexlab{}.
\newblock \showarticletitle{Amazon. com recommendations: Item-to-item
  collaborative filtering}.
\newblock \bibinfo{journal}{\emph{IEEE Internet computing}} \bibinfo{number}{1}
  (\bibinfo{year}{2003}), \bibinfo{pages}{76--80}.
\newblock


\bibitem[\protect\citeauthoryear{Liu, Zeng, Mokhosi, and Zhang}{Liu
  et~al\mbox{.}}{2018}]%
        {liu2018stamp}
\bibfield{author}{\bibinfo{person}{Qiao Liu}, \bibinfo{person}{Yifu Zeng},
  \bibinfo{person}{Refuoe Mokhosi}, {and} \bibinfo{person}{Haibin Zhang}.}
  \bibinfo{year}{2018}\natexlab{}.
\newblock \showarticletitle{STAMP: short-term attention/memory priority model
  for session-based recommendation}. In \bibinfo{booktitle}{\emph{KDD}}. ACM,
  \bibinfo{pages}{1831--1839}.
\newblock


\bibitem[\protect\citeauthoryear{Luong, Pham, and Manning}{Luong
  et~al\mbox{.}}{2015}]%
        {luong2015effective}
\bibfield{author}{\bibinfo{person}{Minh-Thang Luong}, \bibinfo{person}{Hieu
  Pham}, {and} \bibinfo{person}{Christopher~D Manning}.}
  \bibinfo{year}{2015}\natexlab{}.
\newblock \showarticletitle{Effective approaches to attention-based neural
  machine translation}.
\newblock \bibinfo{journal}{\emph{arXiv preprint arXiv:1508.04025}}
  (\bibinfo{year}{2015}).
\newblock


\bibitem[\protect\citeauthoryear{Merity, Keskar, and Socher}{Merity
  et~al\mbox{.}}{2017}]%
        {merity2017regularizing}
\bibfield{author}{\bibinfo{person}{Stephen Merity},
  \bibinfo{person}{Nitish~Shirish Keskar}, {and} \bibinfo{person}{Richard
  Socher}.} \bibinfo{year}{2017}\natexlab{}.
\newblock \showarticletitle{Regularizing and optimizing LSTM language models}.
\newblock \bibinfo{journal}{\emph{arXiv preprint arXiv:1708.02182}}
  (\bibinfo{year}{2017}).
\newblock


\bibitem[\protect\citeauthoryear{Quadrana, Karatzoglou, Hidasi, and
  Cremonesi}{Quadrana et~al\mbox{.}}{2017}]%
        {quadrana2017personalizing}
\bibfield{author}{\bibinfo{person}{Massimo Quadrana},
  \bibinfo{person}{Alexandros Karatzoglou}, \bibinfo{person}{Bal{\'a}zs
  Hidasi}, {and} \bibinfo{person}{Paolo Cremonesi}.}
  \bibinfo{year}{2017}\natexlab{}.
\newblock \showarticletitle{Personalizing session-based recommendations with
  hierarchical recurrent neural networks}. In
  \bibinfo{booktitle}{\emph{RecSys}}. ACM, \bibinfo{pages}{130--137}.
\newblock


\bibitem[\protect\citeauthoryear{Rendle, Freudenthaler, and
  Schmidt-Thieme}{Rendle et~al\mbox{.}}{2010}]%
        {rendle2010factorizing}
\bibfield{author}{\bibinfo{person}{Steffen Rendle}, \bibinfo{person}{Christoph
  Freudenthaler}, {and} \bibinfo{person}{Lars Schmidt-Thieme}.}
  \bibinfo{year}{2010}\natexlab{}.
\newblock \showarticletitle{Factorizing personalized markov chains for
  next-basket recommendation}. In \bibinfo{booktitle}{\emph{WWW}}. ACM,
  \bibinfo{pages}{811--820}.
\newblock


\bibitem[\protect\citeauthoryear{Sarwar, Karypis, Konstan, and Riedl}{Sarwar
  et~al\mbox{.}}{2001}]%
        {sarwar2001item}
\bibfield{author}{\bibinfo{person}{Badrul Sarwar}, \bibinfo{person}{George
  Karypis}, \bibinfo{person}{Joseph Konstan}, {and} \bibinfo{person}{John
  Riedl}.} \bibinfo{year}{2001}\natexlab{}.
\newblock \showarticletitle{Item-based collaborative filtering recommendation
  algorithms}. In \bibinfo{booktitle}{\emph{WWW}}. ACM,
  \bibinfo{pages}{285--295}.
\newblock


\bibitem[\protect\citeauthoryear{Tang, Belletti, Jain, Chen, Beutel, Xu, and
  Chi}{Tang et~al\mbox{.}}{2019}]%
        {tang2019towards}
\bibfield{author}{\bibinfo{person}{Jiaxi Tang}, \bibinfo{person}{Francois
  Belletti}, \bibinfo{person}{Sagar Jain}, \bibinfo{person}{Minmin Chen},
  \bibinfo{person}{Alex Beutel}, \bibinfo{person}{Can Xu}, {and}
  \bibinfo{person}{Ed~H Chi}.} \bibinfo{year}{2019}\natexlab{}.
\newblock \showarticletitle{Towards Neural Mixture Recommender for Long Range
  Dependent User Sequences}.
\newblock \bibinfo{journal}{\emph{arXiv preprint arXiv:1902.08588}}
  (\bibinfo{year}{2019}).
\newblock


\bibitem[\protect\citeauthoryear{Tang and Wang}{Tang and Wang}{2018}]%
        {tang2018personalized}
\bibfield{author}{\bibinfo{person}{Jiaxi Tang} {and} \bibinfo{person}{Ke
  Wang}.} \bibinfo{year}{2018}\natexlab{}.
\newblock \showarticletitle{Personalized top-n sequential recommendation via
  convolutional sequence embedding}. In \bibinfo{booktitle}{\emph{WSDM}}. ACM,
  \bibinfo{pages}{565--573}.
\newblock


\bibitem[\protect\citeauthoryear{Vaswani, Shazeer, Parmar, Uszkoreit, Jones,
  Gomez, Kaiser, and Polosukhin}{Vaswani et~al\mbox{.}}{2017}]%
        {vaswani2017attention}
\bibfield{author}{\bibinfo{person}{Ashish Vaswani}, \bibinfo{person}{Noam
  Shazeer}, \bibinfo{person}{Niki Parmar}, \bibinfo{person}{Jakob Uszkoreit},
  \bibinfo{person}{Llion Jones}, \bibinfo{person}{Aidan~N Gomez},
  \bibinfo{person}{{\L}ukasz Kaiser}, {and} \bibinfo{person}{Illia
  Polosukhin}.} \bibinfo{year}{2017}\natexlab{}.
\newblock \showarticletitle{Attention is all you need}. In
  \bibinfo{booktitle}{\emph{NIPS}}. \bibinfo{pages}{5998--6008}.
\newblock


\bibitem[\protect\citeauthoryear{Wang, Huang, Zhao, Zhang, Zhao, and Lee}{Wang
  et~al\mbox{.}}{2018}]%
        {wang2018billion}
\bibfield{author}{\bibinfo{person}{Jizhe Wang}, \bibinfo{person}{Pipei Huang},
  \bibinfo{person}{Huan Zhao}, \bibinfo{person}{Zhibo Zhang},
  \bibinfo{person}{Binqiang Zhao}, {and} \bibinfo{person}{Dik~Lun Lee}.}
  \bibinfo{year}{2018}\natexlab{}.
\newblock \showarticletitle{Billion-scale Commodity Embedding for E-commerce
  Recommendation in Alibaba}.
\newblock \bibinfo{journal}{\emph{arXiv preprint arXiv:1803.02349}}
  (\bibinfo{year}{2018}).
\newblock


\bibitem[\protect\citeauthoryear{Wang, Guo, Lan, Xu, Wan, and Cheng}{Wang
  et~al\mbox{.}}{2015}]%
        {wang2015learning}
\bibfield{author}{\bibinfo{person}{Pengfei Wang}, \bibinfo{person}{Jiafeng
  Guo}, \bibinfo{person}{Yanyan Lan}, \bibinfo{person}{Jun Xu},
  \bibinfo{person}{Shengxian Wan}, {and} \bibinfo{person}{Xueqi Cheng}.}
  \bibinfo{year}{2015}\natexlab{}.
\newblock \showarticletitle{Learning hierarchical representation model for
  nextbasket recommendation}. In \bibinfo{booktitle}{\emph{SIGIR}}. ACM,
  \bibinfo{pages}{403--412}.
\newblock


\bibitem[\protect\citeauthoryear{Wu, Schuster, Chen, Le, Norouzi, Macherey,
  Krikun, Cao, Gao, Macherey, et~al\mbox{.}}{Wu et~al\mbox{.}}{2016}]%
        {wu2016google}
\bibfield{author}{\bibinfo{person}{Yonghui Wu}, \bibinfo{person}{Mike
  Schuster}, \bibinfo{person}{Zhifeng Chen}, \bibinfo{person}{Quoc~V Le},
  \bibinfo{person}{Mohammad Norouzi}, \bibinfo{person}{Wolfgang Macherey},
  \bibinfo{person}{Maxim Krikun}, \bibinfo{person}{Yuan Cao},
  \bibinfo{person}{Qin Gao}, \bibinfo{person}{Klaus Macherey}, {et~al\mbox{.}}}
  \bibinfo{year}{2016}\natexlab{}.
\newblock \showarticletitle{Google's neural machine translation system:
  Bridging the gap between human and machine translation}.
\newblock \bibinfo{journal}{\emph{arXiv preprint arXiv:1609.08144}}
  (\bibinfo{year}{2016}).
\newblock


\bibitem[\protect\citeauthoryear{Ying, Zhuang, Zhang, Liu, Xu, Xie, Xiong, and
  Wu}{Ying et~al\mbox{.}}{2018b}]%
        {ying2018sequential}
\bibfield{author}{\bibinfo{person}{Haochao Ying}, \bibinfo{person}{Fuzhen
  Zhuang}, \bibinfo{person}{Fuzheng Zhang}, \bibinfo{person}{Yanchi Liu},
  \bibinfo{person}{Guandong Xu}, \bibinfo{person}{Xing Xie},
  \bibinfo{person}{Hui Xiong}, {and} \bibinfo{person}{Jian Wu}.}
  \bibinfo{year}{2018}\natexlab{b}.
\newblock \showarticletitle{Sequential Recommender System based on Hierarchical
  Attention Networks}. In \bibinfo{booktitle}{\emph{IJCAI}}.
\newblock


\bibitem[\protect\citeauthoryear{Ying, He, Chen, Eksombatchai, Hamilton, and
  Leskovec}{Ying et~al\mbox{.}}{2018a}]%
        {ying2018graph}
\bibfield{author}{\bibinfo{person}{Rex Ying}, \bibinfo{person}{Ruining He},
  \bibinfo{person}{Kaifeng Chen}, \bibinfo{person}{Pong Eksombatchai},
  \bibinfo{person}{William~L Hamilton}, {and} \bibinfo{person}{Jure Leskovec}.}
  \bibinfo{year}{2018}\natexlab{a}.
\newblock \showarticletitle{Graph Convolutional Neural Networks for Web-Scale
  Recommender Systems}.
\newblock \bibinfo{journal}{\emph{arXiv preprint arXiv:1806.01973}}
  (\bibinfo{year}{2018}).
\newblock


\bibitem[\protect\citeauthoryear{Yuan, Karatzoglou, Arapakis, Jose, and
  He}{Yuan et~al\mbox{.}}{2019}]%
        {yuan2019simple}
\bibfield{author}{\bibinfo{person}{Fajie Yuan}, \bibinfo{person}{Alexandros
  Karatzoglou}, \bibinfo{person}{Ioannis Arapakis}, \bibinfo{person}{Joemon~M
  Jose}, {and} \bibinfo{person}{Xiangnan He}.} \bibinfo{year}{2019}\natexlab{}.
\newblock \showarticletitle{A Simple Convolutional Generative Network for Next
  Item Recommendation}. In \bibinfo{booktitle}{\emph{WSDM}}. ACM,
  \bibinfo{pages}{582--590}.
\newblock


\bibitem[\protect\citeauthoryear{Zhang, Tay, Yao, and Sun}{Zhang
  et~al\mbox{.}}{2018}]%
        {zhang2018next}
\bibfield{author}{\bibinfo{person}{Shuai Zhang}, \bibinfo{person}{Yi Tay},
  \bibinfo{person}{Lina Yao}, {and} \bibinfo{person}{Aixin Sun}.}
  \bibinfo{year}{2018}\natexlab{}.
\newblock \showarticletitle{Next item recommendation with self-attention}.
\newblock \bibinfo{journal}{\emph{arXiv preprint arXiv:1808.06414}}
  (\bibinfo{year}{2018}).
\newblock


\bibitem[\protect\citeauthoryear{Zhao, Wang, Ye, Gao, Yang, and Chen}{Zhao
  et~al\mbox{.}}{2018}]%
        {zhao2018plastic}
\bibfield{author}{\bibinfo{person}{Wei Zhao}, \bibinfo{person}{Benyou Wang},
  \bibinfo{person}{Jianbo Ye}, \bibinfo{person}{Yongqiang Gao},
  \bibinfo{person}{Min Yang}, {and} \bibinfo{person}{Xiaojun Chen}.}
  \bibinfo{year}{2018}\natexlab{}.
\newblock \showarticletitle{PLASTIC: Prioritize Long and Short-term Information
  in Top-n Recommendation using Adversarial Training}. In
  \bibinfo{booktitle}{\emph{IJCAI}}. \bibinfo{pages}{3676--3682}.
\newblock


\bibitem[\protect\citeauthoryear{Zhu, Li, Zhang, Li, He, Li, and Gai}{Zhu
  et~al\mbox{.}}{2018}]%
        {zhu2018learning}
\bibfield{author}{\bibinfo{person}{Han Zhu}, \bibinfo{person}{Xiang Li},
  \bibinfo{person}{Pengye Zhang}, \bibinfo{person}{Guozheng Li},
  \bibinfo{person}{Jie He}, \bibinfo{person}{Han Li}, {and}
  \bibinfo{person}{Kun Gai}.} \bibinfo{year}{2018}\natexlab{}.
\newblock \showarticletitle{Learning Tree-based Deep Model for Recommender
  Systems}. In \bibinfo{booktitle}{\emph{KDD}}. ACM,
  \bibinfo{pages}{1079--1088}.
\newblock


\end{thebibliography}

\end{document}